\documentclass[aps,twocolumn,showpacs,showkeys,floatfix,longbibliography,superscriptaddress]{revtex4-1}

\usepackage{graphicx}
\usepackage{textcomp}
\usepackage{dcolumn}
\usepackage{amsmath}
\usepackage{amssymb}
\usepackage{epstopdf}
\usepackage{rotating}
\usepackage{color}
\usepackage{natbib}
\usepackage{url}
\setcitestyle{square} 
\begin{document}
 
\title{Longitudinal associations between learning assistants and instructor effectiveness}
 
 
\author{Daniel Caravez}\affiliation{Department of Science Education, California State University Chico, Chico, CA, 95929, USA} 
\author{Angelica De La Torre}\affiliation{Department of Science Education, California State University Chico, Chico, CA, 95929, USA} 
\author{Jayson Nissen}\affiliation{Department of Science Education, California State University Chico, Chico, CA, 95929, USA} 
\author{Ben Van Dusen}\affiliation{Department of Science Education, California State University Chico, Chico, CA, 95929, USA} 
 
\begin{abstract}
A central goal of the Learning Assistant (LA) model is to improve students' learning of science through the transformation of instructor practices. There is minimal existing research on the impact of college physics instructor experiences on their effectiveness. To investigate the association between college introductory physics instructors' experiences with and without LAs and student learning, we drew on data from the Learning About STEM Student Outcomes (LASSO) database. The LASSO database provided us with student-level data (concept inventory scores and demographic data) for 4,365 students and course-level data (instructor experience and course features) for the students' 93 mechanics courses. We performed Hierarchical Multiple Imputation to impute missing data and Hierarchical Linear Modeling to nest students within courses when modeling the associations between instructor experience and student learning. Our models predict that instructors' effectiveness decreases as they gain experience teaching without LAs. However, LA supported environments appear to remediate this decline in effectiveness as instructor effectiveness is maintained while they gain experience teaching with LAs. 
 
 \end{abstract}
\maketitle
\section{Introduction}
A central goal of the Learning Assistant (LA) model is to improve students' learning of science through the transformation of instructor practices \citep{Otero2016LearningNetworks}. Specifically, the LA model aims to leverage the additional in-class and planning supports that LAs provide to increase science instructors' uses of research-based pedagogies. To support LA and faculty collaboration, the LA model trains LAs in pedagogy courses and creates opportunities for instructors and LAs to co-plan for class each week. These structures support instructors to reexamine and refine their pedagogical practices and increase their effectiveness. This study sets out to examine whether instructor's prior experience teaching courses with LAs is associated with improved student learning.
\par 	The impact that K-12 teachers prior experiences have on their effectiveness, as measured by students' learning, has been the focus of significant discussion within the education policy communities \citep{StephenSawchuk2015,Ladd2015,Kini2016}. Researchers have repeatedly found strong links between teacher experience and improved student standardized test scores in math and reading \citep{Ladd2015,Kini2016}. The assumption that teachers' experiences improve their effectiveness has strongly influenced the creation of the K-12 salary schedules and is a significant source of inequity across school settings \citep{Rice2010}. While the impact of K-12 teacher experience on student learning of math and reading has been well documented, the impact of college instructor experience on student learning of physics (or science more generally) has received little attention. 
 
\section{Background}
Increases in K-12 teacher effectiveness vary substantially across educational environments. The general trend within these settings is that teacher experience positively correlates with improved student performance on standardized math and reading tests \citep{Kini2016}. The effects of increasing teachers' experience is strongest over the first several years but continued throughout their careers. Teacher effectiveness was also found to improve most quickly when the teachers were in a supportive environment that facilitated peer collaboration \citep{Kini2016}. The impact of teacher experience on student math test performance was only half as strong in grades 6-8 than in grades 4 and 5 (Rice, 2010). Teacher experience was also associated with larger shifts on tests scores in math than in reading \citep{Kini2016}. Some studies have found that teaching effectiveness can decline over time, particularly in high school settings \citep{Rice2010}. These findings show that teacher experience can be an important factor in student learning, but there is considerable variability in how teacher experience relates to teacher effectiveness. Given the variation in impacts of teacher experience within K-12 educational environments and the substantial differences across K-12 and college systems, it is unclear how predictive the findings are of the impacts of instructor experience in college physics courses. 
 
\par 	In one of the few examinations of physics instructor experience, Pollock and Finkelstein \citep{Pollock2012ImpactsPhysics} tracked eight years of student performance in first and second semester physics courses utilizing Peer Instruction \citep{Mazur1997PeerInstruction} and LAs. They categorized each of the 27 instructors in the study as Inexperienced, Experienced, or Physics Education Researcher (PER) based on their teaching and research backgrounds. Students in classes with Experienced and PER instructors outperformed their peers in classes with Inexperienced instructors. The three instructors initially identified as Inexperienced but who subsequently taught the course multiple times saw significant improvements in their students' performance. The three instructors identified as PER and who taught the course multiple times began with higher student performance but failed to show consistent improvement in student performance as they became more experienced. While the study was exploratory and did not have a control group of instructors not using LAs, it indicated that college physics instructor experiences may impact student performance and that these impacts may vary across instructor groups.  
 
\par	In an examination of the impact of LAs on instructor practices, Otero et al. \citep{Otero2010AModel} found that the 11 instructors interviewed reported that collaborating with LAs on planning lessons was instrumental in increasing their attention to student learning. The investigation, however, did not measure shifts in instructor effectiveness over time. While the support and collaborative aspects of the LA model align with the recommendations made in the K-12 literature \citep{Kini2016}, there are no large-scale studies that have empirically tested the impact of these practices on college physics instructors.  
 
\section{Research Questions}
\par 	To investigate how interactions with LAs relates to physics instructor effectiveness over time, we investigated two questions: 
\begin{enumerate}  
\item What is the impact of instructor experience teaching introductory physics courses on student learning, if any? 
\item Does an instructor's history of using LAs alter the impact of their teaching experience on student learning?
\end{enumerate}

\section{Methods}
The Learning About STEM Student Outcomes (LASSO) platform hosts, administers, scores, and analyzes student pre- and posttest assessments online to provide instructors with objective feedback on their courses. Instructors receive a report on their student performance and can access their students' responses. Data from participating courses are added to the LASSO database where they are anonymized, aggregated with similar courses, and made available to researchers with approved IRB protocols. This study used three years of introductory physics courses data from the LASSO database. We examined data from courses that used the Force Concept Inventory (FCI) \citep{Hestenes1992ForceInventory} or Force and Motion Conceptual Evaluation (FMCE) \citep{Thornton1998Fmce}. We did not differentiate between the FCI and FMCE in the models presented in this paper because our preliminary analysis indicated that doing so did not meaningfully change the model.
 
\par	To clean our data, we removed assessment scores for students if they took less than 5 minutes on the assessment or were less than 80\% complete. We removed entire courses if they had less than 40\% student participation on either the pre- or posttest. After filtering our data was  missing 15\% of the pretest scores and 30\% of the posttest scores. To address missing data, we used Hierarchical Multiple Imputation (HMI) with the hmi and mice packages in R. HMI addresses missing data by (1) imputing each missing data point \emph{m} times to create \emph{m} complete data sets, (2) independently analyzing each data set, and (3) combining the  \emph{m} results using standardized methods \citep{Drechsler2015MultipleSimplicity}. The analysis used 10 imputed datasets. HMI is preferable to using matched data because it maximizes statistical power by using all available data \citep{Drechsler2015MultipleSimplicity}. After cleaning and imputation, our dataset included pre- and posttest scores for 4,365 from 93 courses.
 
\par 	 When setting up a course in LASSO, each instructor answered a pair of questions on the number of times they had previously taught the course with LAs and the number of times they had taught the course without LAs. The majority (83\%) of the instructors reported having prior experience teaching their courses (Figure 1). Many of the instructors reported having always taught the course either with (24\%) or without (32\%) LAs. 
 
\begin{figure}
\includegraphics[width=1\columnwidth]{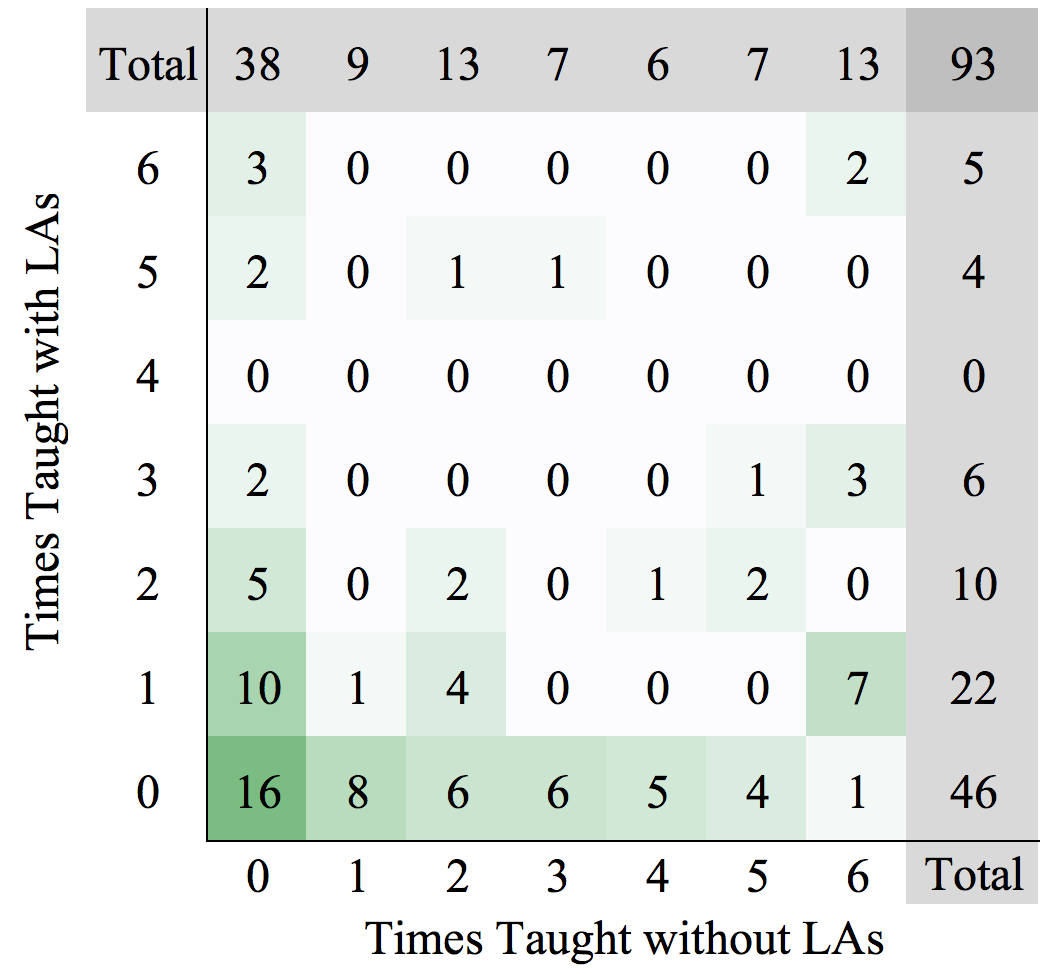}
\caption{Instructor's prior experience teaching their course with and without LAs. Sixteen courses were taught by instructors with no prior experience in that course.}
\end{figure}
 
\par	To identify factors associated with differences in student performance, we developed 2-level HLM models using the HLM 7.01 software. These models nested student data (level 1) within course data (level 2). The multi-level nature of our models allowed us to quantify the impact of teacher experience in courses taught with and without LAs while accounting for inherent and unknown course-level variations (e.g. the time of day of a class and instructor backgrounds, which can lead to unforeseeable differences in student performance).
 
\par	We developed our HLM models through a series of incremental additions of variables. In this paper we show the results from three models. Model 1 is the unconditional model, which predicts the student performance without level-1 or level-2 variables. Model 2 builds on Model 1 by including the student (level-1) variables (student prescore). Model 3 builds on Model 2 by including the course (level-2) variables and is shown below. Postscore is the outcome variable and is in level 1 because it was measured for each student. Level 1 also includes a coefficient for the intercept ($\beta_{0j}$), for the student prescore ($\beta_{1j}$), and for a random effects variable ($r_{0j}$). Each coefficient in level 1 has an associated level 2 equation. In the level 2 equation, the intercept is $\gamma_{i0}$, there is an associated coefficient ($\gamma_{ij}$) for each variable in the equation and $u_{ij}$ represents the random effect for the level 2 equations.
\\ 
\par
\textbf{Level-1 Equation}
\begin{eqnarray*} 
(\text{Postscore})_{ij} & = & \beta_{0j} + \beta_{1j}*(\text{Student Prescore})_{ij} +r_{ij}
 \end{eqnarray*}
 
\textbf{Level-2 Equations}
\begin{eqnarray*}
\beta_{0j}& = & \gamma_{00}+\gamma_{01}*(\text{Times Taught With LAs})_{j} +\\ 
& & \gamma_{02}*(\text{Times Taught Without LAs})_{j}+\\
& & \gamma_{03}*(\text{LA Supported})_{j}+\\
& &  \gamma_{04}*(\text{Class Mean Prescore})_{j}+u_{0j}\\ 
\beta_{1j}& = & \gamma_{10}+u_{1j}\\
\end{eqnarray*}
 
\par 	We do not include LA-support in the level-2 equation for student prescore because the interaction between the variables is not of interest in our analysis. For ease of interpretation, student prescore is group mean centered, class mean prescore is grand mean centered, and all other variables are uncentered. These centerings simplify interpreting the model by shifting the model to predict posttest scores for average performing students in average performing classes. We included pretest scores in the model because they are strong predictors of student performance and improved the model's fit. Pretest scores are not the focus of this investigation so we will not substantively discuss them in our interpretation of the models. 

\par To interpret the models we will first discuss the variance across the three models then we will look at the coefficients for each variable of interest. In discussing the variance, first we will discuss the Intraclass Correlation Coefficient (ICC) for the unconditional model (model 1) to compare the amount of variance in the data at the student versus course level. The ICC is a good indicator of the need for using HLM. Second, we will calculate the variance that the additional variables in each model explained by comparing the variances between the three models. Then we will discuss the model coefficients. 
 

\begin{table}
\caption{Hierarchical Linear Models}
\begin{tabular}{p{2cm}ccp{.05cm}ccp{.05cm}cc}
\hline \hline

\rule{0pt}{3ex}&\multicolumn{8}{c}{Fixed Effects with Robust SE}\\ \cline{2-9}
\rule{0pt}{3ex} &\multicolumn{2}{c}{\underline{Model 1}}&&\multicolumn{2}{c}{\underline{Model 2}}&&\multicolumn{2}{c}{\underline{Model 3}}\\
			&$\beta$&\emph{p}&&$\beta$&\emph{p}&&$\beta$&\emph{p}\\

For Intercept	&&&&&&&&\\
~~Intercept		&55.05&\textless0.001	&&55.04&\textless0.001	&&55.97&\textless0.001	\\
~~With LAs		&-&-&			&-&-&			&0.02&0.974\\
~~W/o LAs	&-&-&			&-&-&			&-1.18&0.01\\
~~LA Sup. 		&-&-&			&-&-&			&3.10&0.171\\
~~Class Pre.	&-&-&			&-&-&			&1.05&\textless0.001\\
\multicolumn{4}{l}{For Student Prescore}	&&&&&\\
~~Intercept		&-&-&		&0.55&\textless0.001	&&0.54&\textless0.001	\\
\hline
\rule{0pt}{3ex}&\multicolumn{8}{c}{Random Effect Variance}\\ \cline{2-9}
Intercept		&\multicolumn{2}{c}{12.81}	&&\multicolumn{2}{c}{12.98}	&&\multicolumn{2}{c}{8.08}	\\
Prescore		&\multicolumn{2}{c}{-}	&&\multicolumn{2}{c}{0.10}	&&\multicolumn{2}{c}{0.10}	\\
Level-1		&\multicolumn{2}{c}{20.14}	&&\multicolumn{2}{c}{17.47}	&&\multicolumn{2}{c}{17.43}	\\

\hline \hline

\end{tabular}
\end{table}

\section{Findings}
The ICC shows that 39\% of the variability in student performance is between students within courses and 61\% of the variability is within students. This shows  that course features are meaningfully related to student performance and HLM is an appropriate method of analysis. The reduction in level-1 variance (\textit{r}) from Model 1 to Model 2 (Table 1) shows that our student-level variable (Student Prescore) explains 13\% of the within-class variance in student performance. The reduction in level-2 intercept variance (\textit{u\textsubscript{0j}}) from Model 1 to Model 3 shows that our final model explains 37\% of the variance in mean performance across classes. This amount of explained variance indicates that Model 3 has strong explanatory power. As Model 3 is our most robust model, we will focus solely on it in the remainder of our findings.
 
\begin{figure}
\includegraphics[width=1\columnwidth]{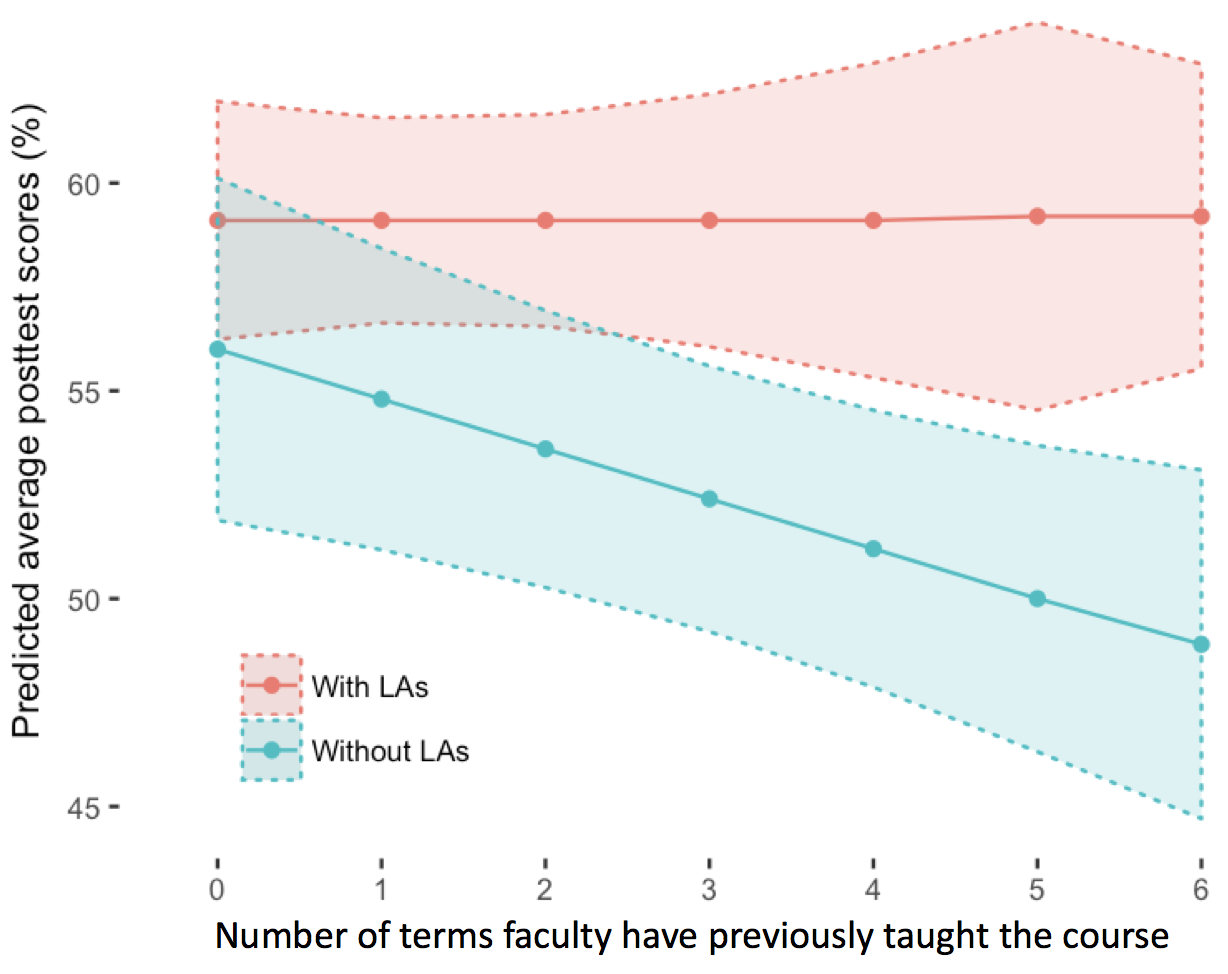}
\caption{Predicted posttest scores for average students with instructors who habitually teach with or without LAs (95\% confidence interval).}
\end{figure}
 
 \par	Model 3 shows that as the number of times instructors have taught a course without LAs increases their posttest scores are predicted to reliably (p=0.01) decrease by 1.2\%/term (Table 1). This decline in instructor effectiveness is absent in LA-supported courses as instructor experience using LAs is not a reliable (p=0.974) predictor of student posttest scores. Figure 2 illustrates the longitudinal impacts of repeated teaching with and without LAs on predicted posttest scores for average students. When instructors have no prior experience teaching that course, our model predicts that the students in LA-supported courses will outperform their peers in courses without LAs by 3.1\%. As the instructors  gain experience the predicted gap in student posttest scores grows. When instructors have 6 terms of experience teaching their respective courses (either with or without LAs), the students in LA-supported courses are predicted to outperform their peers by 10.3\% on their posttests.
\section{Discussion}
Prior research into the impact of experience on teacher effectiveness has primarily focused on K-12 settings. Existing research has found that teacher experience has differential impacts across grades and disciplines. Our investigation extends this work by examining the impact of teaching experience on college introductory physics instructor effectiveness. In answering research question 1, our analysis found a general decline in effectiveness over time for instructors who were not using LAs. In contrast, in answering research question 2, instructors who were teaching with LAs maintained their effectiveness. While the decline in student posttest performance associated with each term of instructor teaching experience without LAs (1.2\%) may not sound substantial, the losses are cumulative. Students in courses with instructors who have taught without LAs for 6 terms are predicted to underperform their peers in LA-supported courses by 10.3\% (7.2\% loss due to instructor experience and 3.1\% loss due to current lack of LA-support). Given that in our dataset the average student raw gains from pretest to posttest are approximately 20\%, losing 10.3\% of that gain represents losing approximately half of the student learning.
\par 	While our models cannot provide a causal mechanism for these observed declines in effectiveness we hypothesize that the shift may come from the lack of resources allocated to improving instructor pedagogical practices. In contrast to K-12 schools, the structures in colleges (e.g. incentives and funding) tend to value faculty's scholarly productivity over teaching effectiveness. Given the general lack of resources that colleges commit to supporting and improving faculty's instructional practices, it is not entirely surprising that college instructors do not improve their effectiveness in the same way as K-12 teachers do. In contrast to typical college instructors, instructors who use LAs have an entire network of resources (e.g. LAs, weekly LA co-planning meetings, LA pedagogy courses, LA program directors, and the LA Alliance) designed to support them in reflecting on and improving their teaching effectiveness. We propose that the additional resources provided by LA programs may help to fill the gap created by the colleges and preventing the decline in instructor effectiveness.

\section {Limitations and future work}
The LASSO platform was originally promoted to faculty in the LA Alliance, which likely skewed the courses to be ones that use research-based pedagogical practices whether they were LA-supported or not. Thus, the shifts in instructor effectiveness without LAs may not represent the shifts in traditional lecture-based courses. We expect that increased adoption of the LASSO platform will improve the generalizability of our findings. In our future work, we will use a meta-analysis of published results to further inform our analysis. This work was funded in part by NSF-IUSE Grant No. DUE-1525338 and is Contribution No. LAA-046 of the Learning Assistant Alliance.

 \bibliography{Mendeley}
 
\end{document}